\documentclass[preprintnumbers,amsmath,amssymb,floatfix,9pt,onecolumn,prd,superscriptaddress,nofootinbib,showpacs,showkeys]{revtex4}
\usepackage{graphicx}
\usepackage{epsfig}
\usepackage{bm}
\usepackage{amsfonts}

\def\ben{\begin{equation}}
\def\een{\end{equation}}

 \def\bd{\begin{document}} \def\ed{\end{document}}
\def\ds{\documentstyle} \let\fr=\frac \let\bl=\bigl \let\br=\bigr
\let\Br=\Bigr \let\Bl=\Bigl
\let\bm=\bibitem
\let\na=\nabla
\let\pa=\partial \let\ov=\overline
\newcommand{\be}{\begin{equation}}
\newcommand{\ee}{\end{equation}}
\def\ba{\begin{array}}
\def\ea{\end{array}}
\def\ft#1#2{{\textstyle{\frac{\scriptstyle #1}{\scriptstyle #2} } }}
\def\fft#1#2{{\frac{#1}{#2}}}
\def\del{\partial}
\def\vp{\varphi}
\def\sst#1{{\scriptscriptstyle #1}}
\def\oneone{\rlap 1\mkern4mu{\rm l}}
\def\td{\tilde}
\def\wtd{\widetilde}
\def\ie{{\it i.e.\ }}
\def\dalemb#1#2{{\vbox{\hrule height .#2pt
        \hbox{\vrule width.#2pt height#1pt \kern#1pt
                \vrule width.#2pt}
        \hrule height.#2pt}}}
\def\square{\mathord{\dalemb{6.8}{7}\hbox{\hskip1pt}}}
\newcommand{\ho}[1]{$\, ^{#1}$}
\newcommand{\hoch}[1]{$\, ^{#1}$}
\newcommand{\bea}{\setlength\arraycolsep{2pt} \begin{eqnarray}}
\newcommand{\eea}{\end{eqnarray}}
\newcommand{\ra}{\rightarrow}
\newcommand{\lra}{\longrightarrow}
\newcommand{\Lra}{\Leftrightarrow}
\newcommand{\bp}{\tilde \beta^\prime}
\newcommand{\tr}{{\rm tr} }
\newcommand{\Tr}{{\rm Tr} }
\def\0{{\sst{(0)}}}
\def\1{{\sst{(1)}}}
\def\2{{\sst{(2)}}}
\def\3{{\sst{(3)}}}
\def\4{{\sst{(4)}}}
\def\5{{\sst{(5)}}}
\def\6{{\sst{(6)}}}
\def\7{{\sst{(7)}}}
\def\8{{\sst{(8)}}}
\def\m{{\sst{(m)}}}
\def\n{{\sst{(n)}}}
\def\cA{{{\cal A}}}
\def\cB{{{\cal B}}}
\def\cF{{{\cal F}}}
\def\cG{{{\cal G}}}
\def\cH{{{\cal H}}}
\def\tV{\widetilde V}
\def\tW{\widetilde W}
\def\tH{\widetilde H}
\def\tE{\widetilde E}
\def\tF{\widetilde F}
\def\tA{\widetilde A}
\def\im{{{\rm i}}}
\def\tY{{{\wtd Y}}}
\def\ep{{\epsilon}}
\def\vep{{\varepsilon}}
\def\bD{{{\bar D}}}
\def\R{{{\mathbb R}}}
\def\C{{{\mathbb C}}}
\def\H{{{\mathbb H}}}
\def\CP{{{\mathbb C}{\mathbb P}}}
\def\RP{{{\mathbb R}{\mathbb P}}}
\def\Z{{{\mathbb Z}}}
\def\bA{{{\mathbb A}}}
\def\bB{{{\mathbb B}}}
\def\bC{{{\mathbb C}}}
\def\bD{{{\mathbb D}}}
\def\bE{{{\mathbb E}}}
\def\bZ{{{\mathbb Z}}}
\def\Re{{{\frak{Re}}}}
\def\Im{{{\frak{Im}}}}
\def\cosec{{\,\hbox{cosec}\,}}
\def\Gm{{\Gamma_{\!\! -}}}
\def\Gp{{\Gamma_{\!\! +}}}
\def\stan{{standard }}
\def\nonstan{{supernumerary }}
\def\p{{\partial}}
\def\kdel#1{{\fft{\del}{\del#1}}}

\def\bog{{Bogomolny }}
\def\om{{\omega}}

\newcommand{\nnr}{\nonumber \\}
\newcommand{\pd}{\partial}
\newcommand{\ud}{\textrm{d}}
\newcommand{\dTH}{T^{\prime \, 0}_\textrm{H}}
\newcommand{\dOi}{\Omega^{\prime \, 0}_i}
\newcommand{\bx}{{\bf x}}
\begin{document}

\title{ Gauss-Bonnet holographic superconductors with magnetic field }

\author{\textbf{M. R. Setare}}
\email{Rezakord@ipm.ir}
 \affiliation{Department of Science, Payame Noor
University, Bijar, Iran }

\author{\textbf{D. Momeni}}
\email{d.momeni@yahoo.com}
 \affiliation{Department of Physics , Faculty of Sciences,
  Tarbiat Moallem University, Tehran, Iran}

\begin{abstract}
We study the Gauss-Bonnet (GB) holographic  superconductors in the
presence of an external magnetic field. We describe the phenomena
away from the probe limit. We derive the critical magnetic field of
the GB holographic superconductors with backreaction. Our analytical
approach matches the numerical calculations.  We calculate the
backreaction corrections up to first order of $O(\kappa^2=8\pi G)$
to the critical temperature $T_C$ and the critical magnetic field
$B_C$ for a GB superconductor. We show that the GB coupling $\alpha$
 makes the condensation weaker but the backreaction corrections $O(\kappa^2)$ make the critical
magnetic field stronger.

\end{abstract}
\pacs{04.70.Bw, 11.25.Tq, 74.20.-z}
 \keywords{Classical Black holes; Gauge/string duality; High-$T_C$ superconductors theory}
 \newpage
 \maketitle

\section{Introduction}
The anti de Sitter/conformal field theory (AdS/CFT) correspondence
\cite{maldacena} provides a powerful theoretical method to
investigate the strongly coupled field theories. It may have useful
applications in condensed matter physics, especially for studying
scale-invariant strongly-coupled systems, for example, low
temperature systems near quantum criticality (see for example
\cite{condencesd} and references therein). Recently, it has been
proposed that the AdS/CFT correspondence also can be used to
describe superconductor phase transition \cite{super}. Since the
high $T_C$ superconductors are shown to be in the strong coupling
regime, one expects that the holographic method could give some
insights into the pairing mechanism in the high $T_C$
superconductors. Inspired by the idea of spontaneous symmetry
breaking in the presence of horizon \cite{gub} various holographic
superconductors have been studied in Einstein theory \cite{GR} or
extended versions as Gauss-Bonnet (GB)\cite{GB1,GB2} ,
Horava-Lifshitz theory \cite{HL1,HL2} and Weyl corrected ones
\cite{weyl}. AdS/CFT can also describe superfluid states in which
the condensing operator is a vector and hence rotational symmetry is
broken, that is, p-wave superfluid states \cite{pwave}. All these
works are based on a numerical analysis of the equations of motion
(EOM) near the horizon and the asymptotic limit by a suitable
shooting method. But as we know that the analytical methods are
better and easy for invoking in different problems. Recently some
attempts have been done on analytical methods in superconductors
(see for example \cite{analytic} and the references in it). In
\cite{analytic} the authors have shown that one can obtain the
critical exponent and the critical temperature by applying a
variational method on the EOM. Their method and terminology is
simple and very sound. Instead of involving in numerical problems ,
we can obtain the critical temperature $T_C$ and the exponent of the
criticality very easily
 by computing a simple variational approach. They studied different
  modes of super criticality
s-wave, p-wave and even d-wave. Their good and efficient method can
be applied on other condensers with higher order Lagrangian black
hole. Recently, we applied this method to superconductors in the
presence of the magnetic field \cite{md}, which our results are in
good agreement with numerical results produced previously
\cite{wen}. Another semi analytical method is based on the matching
method, in which we match the asymptotic solutions at a mid point.
After matching, we can obtain easily the expectation values of the
dual operators $<O_{\pm}>$ and the critical temperature  $T_C$.
Indeed, this method has been used by several authors
\cite{HL2,GB1,kanno}.  In these topics the effect of the external
magnetic field is very important. A holographic model of
superconductor with external magnetic field previously has been
studied numerically \cite{wen}. Here as we can observe, the higher
order terms make the condensation harder. We can study the
backreaction effects in the presence of magnetic field. Our main
goal in this paper is to investigate the effect of the backreaction
on the magnitude of the critical magnetic field $B_C$. We calculate
the backreaction corrections up to first order of $\kappa^2=8\pi G$
to the critical temperature $T_C$ and the critical magnetic field
$B_C$ for a GB superconductor (to see more about $B_C$ in
superconductors refer to \cite{horw}). We show that the GB coupling
$\alpha$ makes the condensation weaker but  the backreaction
corrections make the critical magnetic field stronger.
\section{Gauss-Bonnet holographic superconductors away from the probe
limit}
 We write the action for a Maxwell field and a charged
complex scalar field coupled to the Einstein$-$Gauss$-$Bonnet (EGB)
as
\begin{eqnarray}
S=\frac{1}{2\kappa^2}\int
\sqrt{-g}d^5x[R+12+\frac{\alpha}{2}(R^{\mu\nu\rho\alpha}R_{\mu\nu\rho\alpha}-4R^{\mu\nu}R_{\mu\nu}+R^2)]
\\\nonumber+\int\sqrt{-g}d^5x[-\frac{1}{4}F^{\mu\nu}F_{\mu\nu}-\mid
D_\mu \psi|^2-m^2|\psi|^2],
\end{eqnarray}
where in it the GB coupling is $\alpha$, the AdS radius $l=1$, the
field strength tensor is  defined through $F_{\mu\nu}=\partial_\mu
A_\nu-\partial_\nu A_\mu$, and $D_\mu=\partial_\mu-iq A_\mu$, where
the charge of the scalar field is $q$. The mass of the scalar field
is chosen such that it remains below to the Breitenlohner-Freedman
(BF) bound \cite{bf}.
 For 5-dimension case, the GB coupling
$\alpha$ is bounded within the range $-\frac{7}{36}<\alpha<0.09$
(see for example \cite{hh}). The hairy black hole (BH) solution in
the EGB and with plane symmetry is
\begin{eqnarray}
ds^2=-f(r)e^{-\chi(r)}dt^2+\frac{dr^2}{f(r)}+r^2(dx^2+dy^2+dz^2)
\end{eqnarray}
We choose the gauge $A_\mu=(\phi(r),0,B_C x,0)$, where  $B_C$ is the
critical magnetic field in direction $z$. In the absence of the
magnetic field , the superconductor phase has been described
previously \cite{GB1,GB2}. If we ignore the backreaction, and
setting $\psi=0$ (normal phase), the gravity sector decoupled from
the matter part and the EGB field equations give a charged GB black
hole in AdS background
\begin{eqnarray}
f_0(r)=\frac{r^2}{2\alpha}[1-\sqrt{1-4\alpha(1-\frac{h^4}{r^4})+\frac{8\kappa^2\alpha\rho^2}{3h^2
r^4}(1-\frac{h^2}{r^2})}  ]\\\nonumber \chi_0(r)=0\\\nonumber
\phi_0(r)=\mu-\frac{\rho}{r^2}
\end{eqnarray}
We choose the minus sign of the solutions so that we have a solution
in the Einstein limit of GB theory $\alpha\rightarrow0$. The horizon
locates at $r=h$ and the BH temperature reads as
$T_{BH}=\frac{h}{4\pi}$. The effective asymptotic AdS scale is given
by
\begin{eqnarray}
l_{eff}=\sqrt{\frac{2\alpha}{1-\sqrt{1-4\alpha}}}
\end{eqnarray}
The figure (1) shows the behavior of the $l_{eff}$ for
$-\frac{7}{36}<\alpha<0.09$.

\begin{figure}
\centering
 \includegraphics[width=10cm,angle=0]{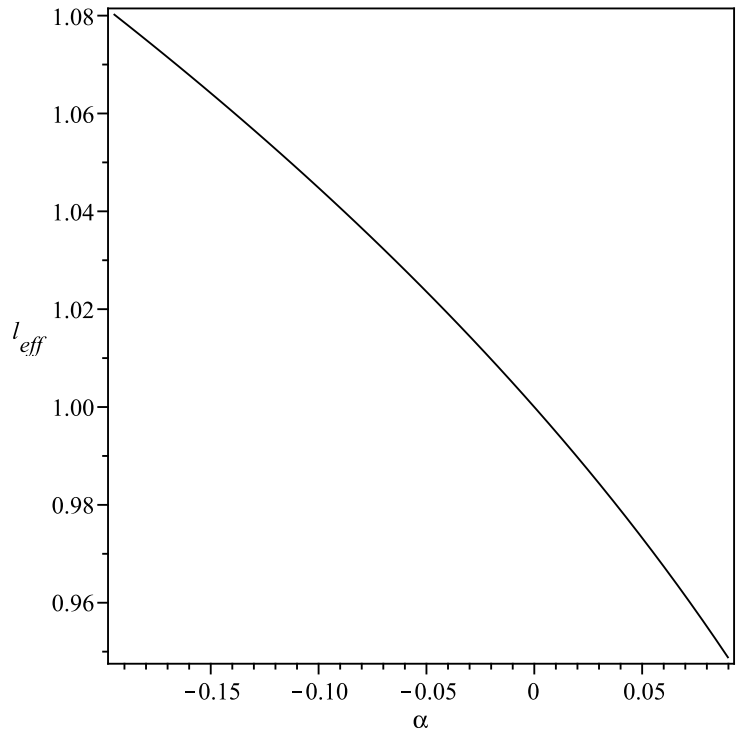}
  \caption{ The $l_{eff}$ as a
function of the coupling  $-\frac{7}{36}<\alpha<0.09$. }
  \label{1.eps}
\end{figure}

 In next section we will use from the
zeroth order solutions given in (3) in a perturbation method based
on the work of \cite{kanno}. The field equations can be written as
the following forms
\begin{eqnarray}
\nabla_\mu F^{\mu\nu}=iq(\psi^{*} D^{\nu}\psi-\psi D^{*\nu}\psi^{*})\\
D_\mu D^\mu\psi-m^2\psi=0\\
R_{\mu\nu}-\frac{1}{2}Rg_{\mu\nu}-6g_{\mu\nu}+4\alpha[R_{\mu\sigma\kappa\tau}R^{\sigma\kappa\tau}_{\nu}\\\nonumber
-2R_{\mu\rho\nu\sigma}R^{\rho\sigma}-2R_{\mu\sigma}R^{\sigma}_{\nu}+RR_{\mu\nu}-\frac{1}{2}
g_{\mu\nu}(R^{\mu\nu\rho\alpha}R_{\mu\nu\rho\alpha}\\\nonumber-4R^{\mu\nu}R_{\mu\nu}+R^2)]=\kappa^2
T_{\mu\nu},
\end{eqnarray}
here the $T_{\mu\nu}$ denotes the total energy-momentum tensor of
matter fields. \\
 In limit of the zero magnetic field $B_C=0$, the
field equations are given by
\begin{eqnarray}
f'=2r\frac{2r^2-f}{r^2-4\alpha f}\\\nonumber-\kappa^2
\frac{r^3e^{\chi}}{f}[\frac{2q^2\phi^2\psi^2+f(2m^2\psi^2
e^{-\chi}+\phi'^{2})+2f^2e^{-\chi}\psi'^{2}}{r^2-4\alpha f}]\\
\chi'e^{-\chi}=-2\kappa^2\frac{r^3(q^2\phi^2 \psi^2+e^{-\chi} f^2
\psi'^{2})}{f^2(r^2-4\alpha f)}\\
\phi''=-(\frac{3}{r}+\frac{\chi'}{2})\phi'+\frac{2q^2\psi^2\phi }{f}\\
\psi''=-(\frac{3}{r}+\frac{f'}{f}-\frac{\chi'}{2})\psi'-(\frac{q^2\phi^2e^{\chi}}{f^2}-\frac{m^2}{f})\psi
\end{eqnarray}
This system has been studied completely using numerical schemas
\cite{GB2}. It has been shown that the critical temperature in the
absence of any magnetic field is $T_C\approx \rho^{1/3}$. For $T <
T_C$ these solutions will be unstable to form scalar hair, i.e.
develop a non-vanishing value of $\psi$ on the horizon. \emph{In the
gauge theory $-$ gravity duality, $ T_C$ is the temperature below
which superconductivity appears}.
\section{Calculating the  corrections to $ B_C$ }
The critical temperature with backreaction has been obtained
analytically in \cite{kanno} and we briefly summarize the main
results first. In order to solve the field equations (8)-(11),
firstly we define a new coordinate $z=\frac{h}{r}$. This map
converts the interval $(h,\infty)$ of the coordinate r to the inside
of the strip $(0,1)$ of the new coordinate z. It is useful in
numerical analysis of the field equations. We examine the near
critical point behavior of the system. It is convenient to introduce
a small parameter for our perturbation analysis
\begin{eqnarray}
\epsilon\equiv<O_{\triangle_{+}}>
\end{eqnarray}
Near the critical point , the value of the scalar field $\psi$ is
small. Thus we can expend the fields , metric functions and the
chemical potential $\mu$ as the following series
\begin{eqnarray}
\phi=\Sigma^{\infty}_{0}\epsilon^{2n}\phi_{2n}\\
\psi=\Sigma^{\infty}_{0}\epsilon^{2n+1}\psi_{2n+1}\\
f=\Sigma^{\infty}_{0}\epsilon^{2n}f_{2n}\\
\nu=\Sigma^{\infty}_{1}\epsilon^{2n}\nu_{2n}\\
\mu=\mu_0+\epsilon^2\delta\mu_2,
\end{eqnarray}
here $\epsilon<<1$. From (17), we obtain the familiar exponent
$\frac{1}{2}$ for phase transition. The critical value of the
chemical potential is $\mu_C=\mu_0$. Including the backreaction
effects on the critical temperature when $\kappa^2<<1$ has been
discussed in \cite{kanno}. The corrected critical temperature in
first order of $\kappa^2$ and with the matching point
$z_m=\frac{1}{2}$  ,mass $m^2=-3$ and the conformal dimension
$\triangle_{+}=3$ is
\begin{eqnarray}
T_C=T_0(1-2\kappa^2 \delta T),
\end{eqnarray}
where in it
\begin{eqnarray}
T_0=\frac{\sqrt[3]{\rho}}{2^{2/3}\pi}(0.778+1.5\alpha)^{-1/6}
\end{eqnarray}
\begin{eqnarray}
\delta T=1.641+2.667\alpha
\end{eqnarray}
The figure 2 shows the behavior of the $T_C/T_0$ as a function of
the GB coupling $\alpha$ for different values of $\kappa^2$. As we
can observe that, when we fix the backreaction coupling, then in
fixed coupling $\kappa^2$, increasing the GB coupling $\alpha$
decreases the rate of $T_C/T_0$. It means that the condensation
becomes harder when we increase the GB coupling in a fixed
backreaction's correction.
\begin{figure}
\centering
 \includegraphics[width=15cm,angle=0]{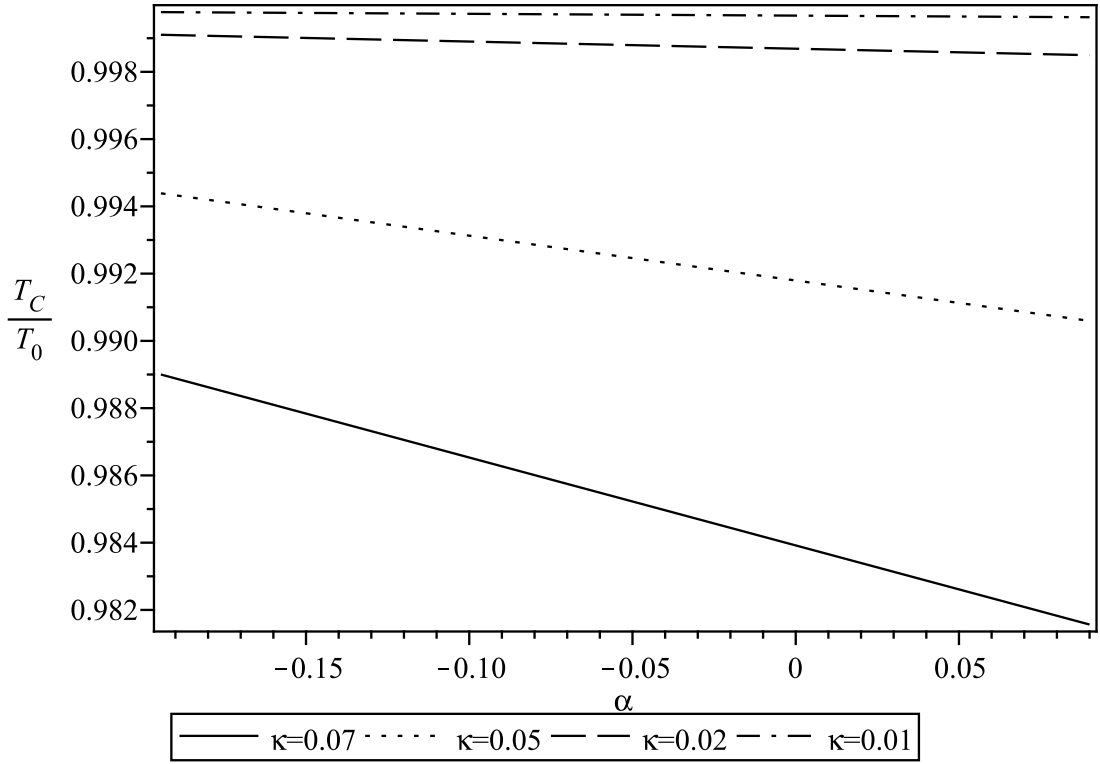}
  \caption{ The ratio of $T_C/T_0$ for a sample of the $\kappa$ as a
function of the coupling  $-\frac{7}{36}<\alpha<0.09$.  }
  \label{2.eps}
\end{figure}

Now, we consider an external magnetic field, so $\psi_1$ satisfy
following differential equation
\begin{eqnarray}
\psi''_1+\bigg(\frac{f'_0}{f_0}-\frac{1}{z}\bigg)\psi'_1+\frac{h^2}{z^4}\bigg(\frac{\phi^2_0
e^{\chi}}{f^2_0}-\frac{m^2}{f_0}\bigg)\psi_1=\\\nonumber-\frac{l^2}{z^2}\bigg[\partial^2_x+(\partial_y-iB_Cx)^2\bigg]\psi_1
\end{eqnarray}

By separation of variables we have

\begin{eqnarray}
\psi_1=e^{ik_y y} X_n(x)Z_n(z),
\end{eqnarray}

where $X(x)$ satisfies following equation for a two dimensional
harmonic oscillator

 \begin{eqnarray} \label{hermite}
-\left(\partial^2_{x} -(k_y-B_Cx)^2\right)X_n(x)={u_{n}B_C}X_n(x).
 \end{eqnarray}
Simply one can find $X(x)$ in terms of  Hermite function $H_n$ as
 \begin{eqnarray}
X(x)=e^{-a(x-x_0)^2} H_n(x),
\end{eqnarray}

where $u_n=2n+1,a=( B_C)/2,x_0=\frac{k_y}{B_C}$. By considering the
lowest mode $n=0$, one can obtain the following equation for
$Z_0(z)$,
 \begin{eqnarray}
Z''_0+\bigg(\frac{f'_0}{f_0}-\frac{1}{z}\bigg)Z'_0+\frac{h^2}{z^4}\bigg(\frac{\phi^2_0
e^{\chi}}{f^2_0}-\frac{m^2}{f_0}\bigg)Z_0=\frac{B_Cl^2}{z^2f_0}Z_0,
\label{Gb}
\end{eqnarray}

 Due to the regularity at the horizon we have

 \begin{eqnarray}
Z'_0=\frac{m^2h^2+B_Cl^2}{f'_0(1)}Z_0(1)
\end{eqnarray}
In details, we must remove the singularity from the term
$\frac{\phi_0^2 e^{\chi}}{f_0^2}$ near the horizon, located at
$z=\frac{h}{r}=1$. From (3), it is obvious that when $z=1$, then
$f_0\rightarrow0,\phi_0^2\rightarrow0$. Using the Hopital's rule for
removing the singularity, from the term $\lim
_{z\rightarrow1}(\frac{h^2}{z^4}\bigg(\frac{\phi^2_0
e^{\chi}}{f^2_0}-\frac{m^2}{f_0}\bigg)-\frac{B_Cl^2}{z^2f_0}))Z_0(z)$,
by simple algebra we obtain the non singular part
$\frac{m^2h^2+B_Cl^2}{f'_0(1)}Z_0(1)$. Collecting it with the same
taking limit process we obtain (26).
 At the AdS boundary $z=0$  we can write

\begin{eqnarray}
Z_0=D_{+}z^{\Delta_{+}},\label{gb}
\end{eqnarray}

where $\Delta_{+}=2+\sqrt{4+m^2l^2_{eff}}$. Now we would like to
obtain the solution of $Z_0$  using the matching method \cite{kanno}
(see also \cite{h}) . For this purpose we can expand $Z_0$ in a
Taylor series near the horizon as

 \begin{eqnarray}
  Z_0=\sum_{n=0}^{\infty}Z^{(n)}_0(1)(1-z)^n
   \end{eqnarray}

 From (\ref{Gb}), we obtain $Z''_0(1)$ as

 \begin{eqnarray}
Z''_0(1)=-\frac{1}{2}\bigg(3+\frac{f''_0(1)}{f'_0(1)}-\\\nonumber\frac{m^2h^2+B_Cl^2}{f'_0(1)}\bigg)Z'_0(1)
-\frac{h^2\phi'_0(1)^2}{2f'^2_0(1)}Z_0(1).
\end{eqnarray}

So an approximate solution near the horizon is given by

 \begin{eqnarray}
Z_0(z)=Z_0(1)-\frac{m^2h^2+B_Cl^2}{f'_0(1)}Z_0(1)(1-z)\\\nonumber-
\bigg[\frac{m^2h^2+B_Cl^2}{4f'_0(1)}\bigg(3+\frac{f''_0(1)}{f'_0(1)}
\\\nonumber-\frac{m^2h^2+B_Cl^2}{f'_0(1)}\bigg)
+\frac{h^2\phi'_0(1)^2}{4f'^2_0(1)}\bigg]Z_0(1)(1-z)^2+O((1-z)^3)\label{gb2}
\end{eqnarray}

Now we connect the solutions Eq.(\ref{gb}) and (30) at the matching
point $z_m=\frac{1}{2}$ smoothly and by defining a new parameter
$\xi=m^2h^2+B_C l^2$ we have

\begin{eqnarray}
(\frac{1}{2})^{\Delta_{+}}D_{+}=Z_0(1)-\frac{\xi}{2f'_0(1)}Z_0(1)
\\\nonumber-\bigg[\frac{\xi}{4f'_0(1)}\bigg(3+\frac{f''_0(1)}{f'_0(1)}
-\frac{\xi}{f'_0(1)}\bigg)
+\frac{\xi}{4f'^2_0(1)}\bigg]\frac{Z_0(1)}{4},\label{c1}\\
\Delta_{+}(\frac{1}{2})^{\Delta_{+}-1}D_{+}=\frac{\xi}{f'_0(1)}Z_0(1)
\\\nonumber+\bigg[\frac{\xi}{4f'_0(1)}\bigg(3+\frac{f''_0(1)}{f'_0(1)}
-\frac{\xi}{f'_0(1)}\bigg)
+\frac{h^2\phi'_0(1)^2}{4f'^2_0(1)}\bigg]\frac{Z_0(1)}{2}.
\end{eqnarray}

 From these equations, we obtain the relation between  $D_{+}$,
 and $Z_0(1)$

\begin{eqnarray}
 D_{+}=\frac{2^{\Delta_{+}}}{1+\frac{\Delta_{+}}{2}}
\bigg(1-\frac{\xi}{4f'_0(1)}\bigg)Z_0(1),\label{dp}
\end{eqnarray}

Substituting $D_{+}$ from above equation into Eq. (30), we find
following relation in case of $Z_0(1)\neq 0$

\begin{eqnarray}
 \frac{2\Delta_{+}}{1+\frac{\Delta_{+}}{2}}
-\bigg(\frac{\frac{\Delta_{+}}{2}}{1+\frac{\Delta_{+}}{2}}+\frac{7}{4}\bigg)
\frac{\xi}{f'_0 (1)} -\frac{\xi}{4}\frac{f''_0(1)}{{f'_0
(1)}^2}\\\nonumber+\frac{1}{4}\frac{\xi^2}{{f'_0 (1)}^2}
-\frac{h^2}{4}\frac{\phi'_0(1)^2}{{f'_0 (1)}^2}=0.\label{relation}
\end{eqnarray}

By substituting the  values of $f'_0(1)$, $f''_0(1)$ and
$\phi'_0(1)$ into (\ref{relation}), we obtain an equation for
$\mu_0$

 \begin{eqnarray}
 \frac{\kappa^4
l^4}{9h^4}\bigg[\frac{2\Delta_{+}}{1+
\frac{\Delta_{+}}{2}}-\frac{l^2\xi\alpha}{2h^2}\bigg]\mu^4_0
\\\nonumber-\frac{l^4}{16h^2}\bigg\{1+2\kappa^2
\bigg[\bigg(\frac{32}{3l^2}+\frac{\xi
l^2}{3h^2}\bigg)\frac{\Delta_{+}}
{1+\Delta_{+}/2}\nonumber\\+\frac{2\xi}{h^2}
-\frac{8\xi}{3h^2}\alpha]\}\mu_0^2+\frac{
\Delta_{+}/2}{1+\frac{\Delta_{+}}{2}}\bigg(4+
\frac{l^2\xi}{4h^2}\bigg)+\frac{3}{8} \frac{\xi^2
l^4}{h^4}\frac{1}{64} -\frac{\xi l^2}{h^2}\alpha=0.
\end{eqnarray}

In the above equation we assume that $\kappa^4 \ll 1$, then by
considering the relation $\mu_0=\frac{\rho}{h^2}$ (which is obtained
from $\phi_0(1)=0$ from Eq. (2)) , we can write following expression
for $B_C$.

 \begin{eqnarray}
 B_C=B_1+\kappa^2\delta B,
\end{eqnarray}

 The resulting critical magnetic
field is the upper critical magnetic field, not the lower one .
 \begin{eqnarray}
B_1=\frac{\pi^2
T^2}{(2+\Delta_{+})}\bigg\{2\bigg[16(3-4\alpha)^2\\\nonumber+16(7-32\alpha+16\alpha^2)\Delta_{+}
+4(16\alpha^2-40\alpha+9)\Delta^2_{+}
\\\nonumber+\frac{(2+\Delta_{+})(96\alpha
\Delta_{+}+192\alpha+145\Delta_{+}
-126)T^6_c}{4T^6}\bigg]^{1/2}\\\nonumber-(18+17\Delta_{+}-32\alpha-16\Delta_{+}\alpha)\bigg\}\\
\delta B=\frac{4\pi^2 T^6_c}{3T^4(2+\Delta_{+})^2}(96\alpha
\Delta_{+}+192\alpha+145\Delta_{+}-126)\bigg\{3+2\Delta_{+}
-2\alpha(2+\Delta_{+})\\\nonumber-2\bigg[18+19\Delta_{+}+6\Delta^2_{+}
+8\alpha^2(2+\Delta_{+})^2-6\alpha(8+10\Delta_{+}+3\Delta^2_{+})\bigg]\bigg/\bigg[16(3-4\alpha)^2\\\nonumber
+16(7-32\alpha+16\alpha^2)\Delta_{+}+4(16\alpha^2-40\alpha+9)\Delta^2_{+}
\\\nonumber+\frac{(2+\Delta_{+})(96\alpha
\Delta_{+}+192\alpha+145\Delta_{+}-126)T^6_c}{4T^6}\bigg]^{1/2}\bigg\}
\end{eqnarray}

\begin{figure}
\centering
 \includegraphics[width=15cm,angle=0]  {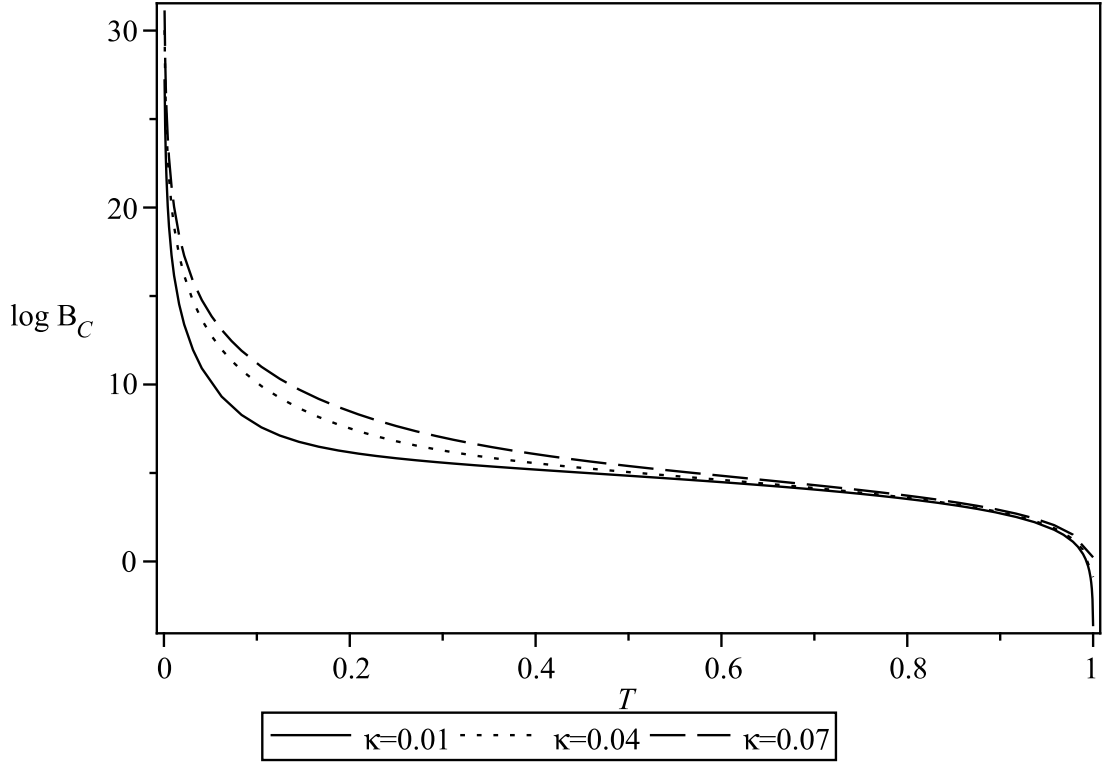}
\caption{ The plot of the Critical magnetic field $ \log B_C$  as a
function of the temperature $T$ for different values of the
backreaction $\kappa$ for GB coupling $\alpha=0.01$.}
 \label{3.eps}
\end{figure}

\begin{figure}
\centering
 \includegraphics[width=15cm,angle=0]  {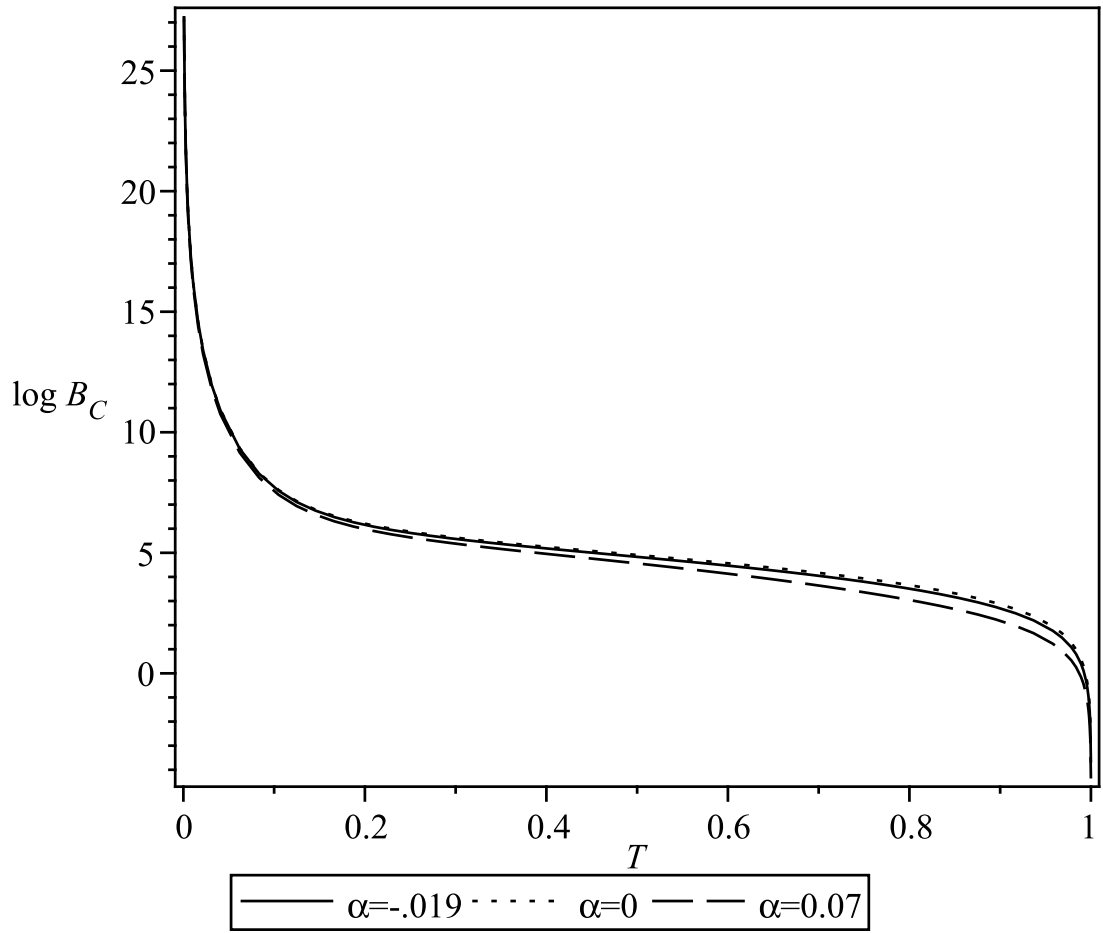}
\caption{ The plot of the Critical magnetic field $ \log B_C$  as a
function of the temperature $T$ for different values of the GB
coupling  $\alpha$ for the value of   $\kappa=0.01$.}
 \label{4.eps}
\end{figure}
In the next figures (3,4), we plot the variation of the $ \log B_C$
as a function of the temperature $T/T_c$ for different values of the
backreaction $\kappa$, GB coupling $\alpha$. The figure (3) shows
the variation of $ \log B_C$ for $\alpha=0.01$. As we observe that
when we fix the GB coupling at $\alpha=0.01$, varying the
backreaction effects term $\kappa^2$, the critical magnetic field,
with backreactions corrections becomes larger. Thus the backreaction
terms $\kappa^2$ in fixed GB coupling $\alpha$ makes the
condensation harder. It is observed from figure (4), in which we
fixed the backreaction effects term $\kappa=0.01$ and varying it
with respect to the GB coupling $\alpha=-0.19,0,0.07$,the effect of
changes in $\alpha$ makes the critical magnetic field weaker,
because it is always smaller than the critical magnetic field to
$\alpha=0$.
 As one can see $\delta B$ is
positive for possible values of the Hawking temperature below the
critical temperature and the Gauss-Bonnet coupling. Due to this
backreaction makes the critical magnetic field stronger. Previously
such investigation in the case $\kappa^2=0$ have been done by Ge,
et. al in \cite{gw}. Our result in this special case is match with
the result of \cite{gw}. One can see the behavior of backreaction
term $\delta B$ with respect to temperature ,it shows that higher
temperature $T$ leads to a lower $\delta B$.

\section{Conclusions}
There are many interesting features for critical phenomena and
superconductivity when we are working on higher orders corrections,
specially when we are interesting in the Gauss- Bonnet corrections
\cite{GB1,GB2}. In the present paper we have investigated the
implication of Gauss-Bonnet correction to the holographic
superconductor in the presence of an external magnetic field, and
away from probe limit. Here we have done our calculation
analytically, where the results are matches with the numerical
calculations. We have obtained the critical magnetic filed $\log
B_C$ up to order $k^2$. The resulting critical magnetic field is the
upper critical magnetic field, not the lower one . Our results show
that the GB coupling $\alpha$  makes the critical magnetic field
weaker, because it is always smaller than the critical magnetic
field to $\alpha=0$. But the backreaction corrections make the
critical magnetic field stronger.
\section{Acknowledgement}The authors would like to thank X. H. Ge, Bin Wang, S. F. Wu and G. H. Yang
 for reading the manuscript.
\section{Note added}When we was busy with the calculation for this
paper, a paper \cite{h}  appeared in arXiv where the similar problem
have been discussed.


\begin{thebibliography}{99}




\bibitem{maldacena} J. M. Maldacena, Adv. Theor. Math. Phys. 2, 231 (1998) [Int. J. Theor. Phys. 38, 1113 (1999)]
[arXiv:hep-th/9711200].
\bibitem{condencesd}
C. P. Herzog, J. Phys. A 42 (2009) 343001, arXiv:0904.1975. ; S. A.
Hartnoll, Class. Quant. Grav. 26 (2009) 224002, arXiv:0903.3246.
\bibitem{super}
G. Policastro, D. T. Son, and A. O. Starinets, Phys. Rev. Lett. 87,
081601 (2001). ; P. Kovtun, D. T. Son, and A. O. Starinets, JHEP10
(2003) 064.
\bibitem{gub}S. S. Gubser, Class. Quant. Grav. 22, 5121 (2005); S. S. Gubser, Phys. Rev. D 78, 065034
(2008).
\bibitem{GR}S. A. Hartnoll, C. P. Herzog, and G. T. Horowitz, Phys. Rev. Lett. 101, 031601
(2008); S. A. Hartnoll, C. P. Herzog, and G. T. Horowitz, J. High
Energy Phys. 12 (2008) 015 ; G. T. Horowitz and M. M. Roberts, Phys.
Rev. D 78, 126008 (2008).
\bibitem{GB1}
R. Gregory, S. Kanno and J. Soda, J. High Energy Phys. 0910 (2009)
010 [arXiv:0907.3203 [hep-th]].

\bibitem{GB2}Q. Pan, B. Wang, E. Papantonopoulos, J. de Oliveira, A. B. Pavan
 Phys. Rev. D81, 106007, (2010);
 Y. Brihaye and B. Hartmann, Phys. Rev.
D 81, 126008 (2010).; R. G. Cai, Z. Y. Nie, and H. Q. Zhang, Phys.
Rev. D 82, 066007 (2010); Q. Pan , B. Wan, Phys. Lett. B 693, 159,
(2010); Q. Pan, J. Jing, B. Wang, arXiv:1105.6153 [gr-qc].
 .

\bibitem{HL1}
R. G. Cai, H. Q. Zhang, Phys. Rev. D81, 066003, (2010).
\bibitem{HL2}
 Davood Momeni, M. R. Setare, N. Majd, J. High Energy
Phys. 05 (2011) 118, arXiv:1003.0376
\bibitem{pwave}
S. S. Gubser, S. S. Pufu, J. High Energy Phys. 0811 (2008) 033,
arXiv:0805.2960; C. P. Herzog, S. S. Pufu, J. High Energy Phys. 0904
(2009) 126, arXiv:0902.0409; P. Basu, J. He, A. Mukherjee, H. H.
Shieh, Phys. Lett. B689, 45, (2010);
 M. Ammon et al., Phys Lett. B 686, 192, (2010).

\bibitem{analytic}
H. B. Zeng, X. Gao, Y. Jiang, H. Sh. Zong, J. High Energy Phys. 05
(2011) 2,arXiv:1012.5564 [hep-th].

\bibitem{weyl}
J. P. Wu, Y. Cao, X. M. Kuang, W.J. Li, Phys. Lett. B697, 153,
(2011), [arXiv:1010.1929v3 [hep-th]];Davood Momeni, M.R. Setare,
arXiv:1106.0431.
\bibitem{md}D. Momeni, Eiji Nakano, M. R. Setare and Wen-Yu Wen, arXiv:1108.4340v1
[hep-th].
\bibitem{wen}E. Nakano and Wen-Yu Wen, Phys. Rev. D 78, 046004
(2008); T. Albash and C. V. Johnson, J. High Energy Phys. 0809, 121
(2008) [arXiv:0804.3466 [hep-th]].
\bibitem{bf}
P. Breitenlohner and D. Z. Freedman, "Positive Energy in anti-De
Sitter Backgrounds and Gauged Extended Supergravity", Phys. Lett.
B115 (1982) 197.
\bibitem{hh}X.-H. Ge, Y. Matsuo, F.-W. Shu,  S.-J. Sin,  T. Tsukioka,
JHEP 0810, 009, (2008); X. -H. Ge, S.-J. Sin, J. High Energy Phys.
0905, 051, (2009).

\bibitem{kanno}
S. Kanno, Class. Quant. Grav.28 (2011) 127001
[arXiv:1103.5022[hep-th]].
 \bibitem{horw}G. T. Horowitz, arXiv:1002.1722.

\bibitem{gw}
 X. H. Ge, B. Wang, S. F. Wu and G. H. Yang,
J. High Energy Phys. {\bf 1008} (2010) 108 [arXiv:1002.4901
[hep-th]]

\bibitem{h}X. H. Ge, "Analytical calculation on critical magnetic field
in holographic superconductors with backreaction", arXiv:1105.4333
[hep-th].







\end{thebibliography}
\end{document}